\documentclass{article}
\usepackage{amsfonts}
\usepackage{amsmath}
\usepackage[utf8]{inputenc}
\usepackage{subcaption}
\usepackage{graphicx}
\title{Machine Learning Symmetry}
\author{Shailesh Lal \footnote{shailesh.hri@gmail.com}}

\date{}
\begin{document}
\maketitle

\centerline{\it Faculdade de Ciencias, Universidade do Porto}
\centerline{\it 687 Rua do Campo Alegre, Porto, Portugal}

\begin{abstract}
    We review recent work in machine learning aspects of conformal field theory 
    and Lie algebra representation theory using neural networks.
\end{abstract}
\section{Introduction}
One of the most striking examples of the success of machine learning and neural networks 
in analyzing physical systems has been in the context of the string landscape problem, initiated in \cite{He:2017aed,Krefl:2017yox,Ruehle:2017mzq,Carifio:2017bov}, see \cite{He:2020mgx, Ruehle:2020jrk} for reviews. Neural networks have been used with great accuracy to predict various properties of compactification manifolds such as topological numbers \cite{He:2017set,Erbin:2020srm,Erbin:2020tks,Erbin:2021hmx}, 
bundle cohomology \cite{Ruehle:2017mzq,Klaewer:2018sfl,Brodie:2019dfx}, metrics 
\cite{Ashmore:2019wzb,Larfors:2021pbb}, and symmetry properties \cite{Gao:2021xbs}. 
Machine Learning has also proved useful
to find patterns in particle masses \cite{Gal:2020dyc} and more generally to statistical predictions and model-building in string theory \cite{Deen:2020dlf,Halverson:2019tkf,Halverson:2020opj,Larfors:2020ugo}. Interestingly, neural networks have been also used to even quantify the degree of likelihood that two string vacua will give rise to similar phenomenology \cite{He:2021eiu}. At the heart of these successes is the robust pattern-finding ability of neural networks which allow us to directly infer the output given an input by learning heuristic association rules from previously available data.

If we, somewhat reductively, 
pose the landscape problem as the problem of scanning across a large
parameter space to search for points where specific properties are realized, coupled with
the computational complexity of probing even a single point, then
it is apparent that the landscape 
problem is not unique to string theory but is quite generic to mathematical physics. For 
example, one might ask: `among which in the set of representations of $\mathfrak{su}(5)$
may one embed the matter content of the standard model'? Or equally well, `where in the set of conformal field theories does a given OPE coefficient take a particular value'? For an individual $\mathfrak{su}(5)$ representation or an individual conformal field theory, the answer may
be obtained through a direct, albeit possibly involved, computation. We however, would like to address these questions across an in principle infinite set of possibilities. Doing so might unveil new structures and organizing
principles for the given parameter space, and eventually to new insight.
In the following, we will report on some initial progress in using neural networks for these purposes \cite{Chen:2020dxg,Chen:2020jjw}.

The works \cite{Chen:2020dxg,Chen:2020jjw} are in a sense 
part of a more general program of applying
methods from data science and machine learning to study problems in mathematical physics,
see \cite{He:2018jtw,He:2021oav} for a pedagogical treatment.
By now, machine learning and data science have been used to learn various aspects of
number theory \cite{Alessandretti:2019jbs,He:2020eva,He:2020kzg}, quiver gauge theories and
cluster algebras \cite{Bao:2020nbi}, knot theory \cite{Jejjala:2019kio,Gukov:2020qaj,natureknot} as well
as graph theory \cite{He:2020fdg}
and commutative algebra \cite{Amoros:2021rrx}. Importantly, and in hindsight somewhat as a precursor of
\cite{Chen:2020jjw}, support vector machines and neural networks were used to learn
the algebraic structures of discrete groups and rings \cite{He:2019nzx}.  It was explicitly
shown there that these machine learning frameworks are able to solve difficult problems in the representation theory of discrete groups without having to appeal to computationally expensive methods like Sylow theorems. In addition, neural network regressors have also been trained to identify Lie algebra symmetries in datasets in \cite{Craven:2021ems}.
\section{Machine Learning Conformal Field Theory}
Consider $d$-dimensional Euclidean space with the inner product
\begin{equation}
\vec{x}^2 = x_1^2 + x_2^2 +\ldots+x_d^2
\end{equation}
and the symmetry transformations
\begin{equation}
    \vec x\mapsto \vec x+\vec c,\,\vec x\mapsto R\cdot \vec x,\,\vec x\mapsto \lambda \vec x,\,\vec x\mapsto \frac{\vec x}{\vert\vec{x}\vert^2}\,,
\end{equation}
A conformal field theory, or CFT, is a quantum field theory which is invariant
under these symmetries. These theories
are in ubiquitous physics, being the formalism to describe systems as diverse as phase transitions in a table-top experiment to quantum gravity. 
The past few years have seen a rapid spurt of new numerical and analytical results for
CFT by means of the conformal bootstrap. In particular, one starts with a few minimal
assumptions about a putative CFT, such as the spectrum of light operators.
Then crossing symmetry of four-point functions, along with unitarity, strongly constrain
the possible dynamics of such theories, see \cite{Poland:2018epd} for a review. These developments have opened up the possibility of exploring the landscape of these theories by robust numerical methods. 

It further 
turns out that various defining properties of Conformal Field Theory are machine 
learnable to a near hundred percent accuracy \cite{Chen:2020dxg}. We will focus here on
the conformal block expansion of four point correlation functions of scalar 
conformal primaries. These are operators that transform under conformal transformations as
\begin{equation}
    \mathcal{O}_{\Delta}\left(\lambda\,x\right) = \lambda^{-\Delta}\,\mathcal{O}_{\Delta}\left(\lambda\,x\right)\,.
\end{equation}
Here $\Delta$ is known as the conformal dimension of the operator $\mathcal{O}_\Delta$.
In the following, we will set $d=1$. The
4-point function of the operator $\mathcal{O}_\Delta$ may then be expressed as \cite{Hogervorst:2017sfd}
\begin{equation}
\left\langle \mathcal{O}_\Delta\left(x_1\right) \mathcal{O}_\Delta\left(x_2\right)
\mathcal{O}_\Delta\left(x_3\right) \mathcal{O}_\Delta\left(x_4\right) \right\rangle
=\frac{\sum_{\lbrace h\rbrace} c_h\,^2\,z^h\,_{2}F_1(h,h,2h; z)}{\vert x_{12}\vert^{2\Delta} \vert x_{34}\vert^{2\Delta}}\,.
\end{equation}
Here $z$ is the cross ratio
\begin{equation}
z = \frac{\vert x_{12}\vert \vert x_{34}\vert }{\vert x_{13}\vert \vert x_{24}\vert}\,,
\end{equation}
and the sum runs over all operators $\mathcal{O}_h$ that appear in the OPE $\mathcal{O}_\Delta\times\mathcal{O}_\Delta$, and $c_h$ is the corresponding OPE coefficient.
Here the conformal dimention of $\mathcal{O}_h$ is $h$ and the functions $z^h\,_{2}F_1(h,h,2h; z)$ are the conformal blocks.

\begin{figure}[]
  \centering
  \subcaptionbox{Functions with $c_n \neq 0$.\label{fig:allblock}}{\includegraphics[width=0.45\textwidth]{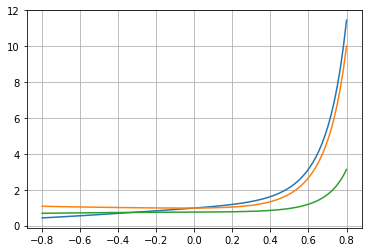}}
  \hspace{1em}%
  \subcaptionbox{Functions with $c_2 = 0$.\label{fig:missingblock}}{\includegraphics[width=0.45\textwidth]{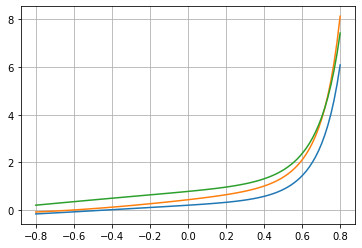}}
  \caption{Toy four-point functions for CFT$_1$.}
  \label{fig:toy4pt}
\end{figure}

In the following, we will demonstrate how neural networks distinguish between two classes
of four-point functions, namely ones where a particular coefficient $c_h$ are zero and 
non-zero respectively. To do so, it is useful to define an auxiliary function
\begin{equation}
    f\left(z\right) = \sum_{\lbrace h\rbrace} c_h\,^2 z^h\,_{2}F_1(h,h,2h; z)\,.
\end{equation}
The spectrum of conformal dimensions $\lbrace h\rbrace$ and
the OPE coefficients $c_h$ are not arbitrary and are constrained so that the overall 
four-point function is crossing symmetric.
We will however consider a toy setting where the $h$ are fixed to be positive integers and
the $c_h$ are generated randomly. 
We then define 
two classes of functions, \textit{viz.}
\begin{equation}\label{eq:f1f2}
    f_1\left(z\right) = \sum_{n=0}^{10} c_n\,^2z^n\,_{2}F_1(n,n,2n; z)\,,
    \quad
    f_2\left(z\right) = \sum_{n=0,\, n\neq 2}^{10} c_n\,^2z^n\,_{2}F_1(n,n,2n; z)\,,
\end{equation}
Some examples of these two classes of functions are visualized in Figures \ref{fig:allblock} and \ref{fig:missingblock}.
To prepare this problem for machine learning, we discretize $z$ to lie on a grid of 100
equidistant points in $\vert z \vert \leq 0.8$ and 
generate 100000 instances each of functions
$f_1$ and $f_2$ by randomly sampling the $c_h$ in the interval $(0,1)$. A relatively 
simple $\mathtt{relu}$ activated MLP architecture trained on this data achieves near perfect performance \cite{Chen:2020dxg}. In particular, when evaluating on 2000 test functions, we obtain the confusion matrix
\begin{equation}
\begin{pmatrix} 1000 & 0 \\ 67 & 933 \end{pmatrix}\,.
\end{equation}
In sum, simple MLP architectures are able to distinguish between two classes of functions
which look quite similar at first glance. In fact, it is possible to understand this difference quite explicitly. Principal component analysis applied to the randomly
generated point clouds of Equation \ref{eq:f1f2} clearly visualizes the two classes of 
functions as being separated from each other.
\begin{figure}[!h]
  \centering
      \includegraphics[width=0.8\linewidth]{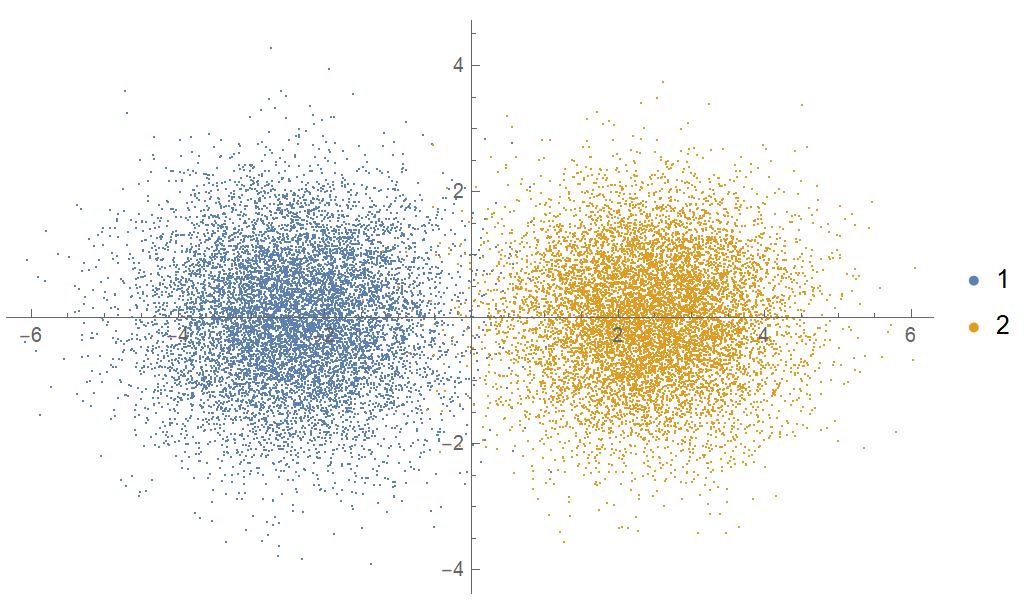}
  \caption{The 
  point clouds corresponding to $f_1$, shown as blue, and $f_2$, shown as yellow, represented in two dimensions by principal component analysis. Each point corresponds to a particular function $f$.}
  \label{fig:pca}
\end{figure}
\section{Machine Learning Lie Algebras}
One of the most fundamental problems in physics is the description of symmetries of a system, i.e. the set of operations that leaves the system invariant. This set has a natural group structure that may be imposed on it. The composition of two symmetry operations is quite clearly a symmetry operation, and any symmetry operation can also be inverted to restore the system to its original state. We will focus on the groups which have elements continuously connected to the identity, i.e. these elements may be parametrized as
\begin{equation}
    U\left(\theta\right) = \exp\left(i\theta^a\,T_a\right)\,, \theta^a\in\mathbb{R}\,.
\end{equation}
These are the Lie groups, whose generators $T^a$ form a Lie algebra, i.e. they are 
endowed with a composition rule $[\cdot,\cdot]$ known as the Lie bracket that obeys, among other properties,
\begin{equation}
    \left[T_a,T_b\right] = i\,f_{ab}\,^c\,T_c\,.
\end{equation}
Physicists usually focus on \textit{unitary representations} of the group $U$ 
over a vector space $\mathcal{V}$. The $T_a$ then are represented by hermitian matrices on $\mathcal{V}$, and the Lie bracket is the usual matrix commutator. 

A representation for which all the $T_a$ cannot be simultaneously brought to block diagonal form is called \textit{irreducible}. Representations which are not irreducible are called \textit{reducible}, and may be expressed as a direct sum of irreducible representations. If a subset of the Lie algebra closes within itself under the same Lie bracket, it is known as a subalgebra. Further, irreducible  representations of a Lie algebra decompose in general into a direct sum of irreducible representations of a given subalgebra. Carrying out such decompositions is in general a complicated analytical task \cite{Slansky:1981yr,Feger:2012bs}. We refer the reader to \cite{Fuchs:1997jv} for more details regarding Lie algebras and their representation theory.

Given that particles carry unitary irreducible representations of the symmetry group, see
e.g. \cite{weinberg}, 
such computations are of key relevance to particle physics model building. As an example, 
consider the Lie algebras $\mathfrak{su}(N)$. These are the algebras corresponding to the group of unitary
$N\times N$ matrices of determinant 1. The
algebra $\mathfrak{su}(5)$ contains as a subalgebra $\mathfrak{su}(3)\times \mathfrak{su}(2) \times \mathfrak{u}(1)$, the symmetry algebra of the standard model. Classifying which representations of $\mathfrak{su}(5)$ contain the 
matter content of the standard model is an important problem for particle physics model building. Such computations are seen to run on exponential time scales \cite{Chen:2020jjw} when carried out on \texttt{LieART}, a \texttt{Mathematica} software for computations in Lie algebra representation theory \cite{Feger:2012bs}. 

To machine-learn this decomposition, we note that unitary irreducible representations of $\mathfrak{su}(5)$ are labeled by an ordered $4$-tuple of integers, see e.g. \cite{Fuchs:1997jv}. We restrict ourselves to those $4$-tuples for which each individual entry ranges from 0 to 4. The dimensions of these representations range from 1, for $(0,0,0,0)$ to 9765625 for $(4,4,4,4)$.
\begin{figure}
  \centering
  \begin{subfigure}[b]{0.45\linewidth}
    \includegraphics[width=\linewidth]{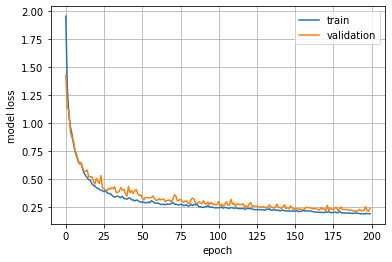}
    \caption{Accuracy}
  \end{subfigure}
  \begin{subfigure}[b]{0.45\linewidth}
    \includegraphics[width=\linewidth]{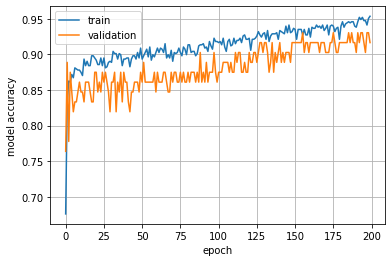}
    \caption{Loss}
  \end{subfigure}
  \caption{Training curves for $\mathfrak{su}(5)$ decomposition.}
  \label{fig:su5}
\end{figure}
This data is again fed to a \texttt{relu} activated MLP \cite{Chen:2020jjw}.
The training curves for this classification are shown in Figure \ref{fig:su5}. The neural network trains to 90\% test accuracy with a Matthews' phi-coefficient of 0.81. 
The confusion matrix obtained is
\begin{equation}
\begin{pmatrix}
43 &  0\\  8 & 29
\end{pmatrix}\,.
\end{equation}
\section{Outlook}
In summary we see that neural networks are able to learn important facts about mathematical physics landscapes, both for conformal field theory and for the representation theory of Lie algebras. We hope that these promising results pave the way for a more exhaustive exploration of these landscapes using methods from data science and machine learning. 
We note especially the recent works on two dimensional CFTs using 
classifiers as well as autoencoders \cite{Kuo:2021lvu}, along with the application of reinforcement learning to solve the equations of the conformal bootstrap itself \cite{Kantor:2021jpz}.
It would also be interesting to examine 
if the methods of \cite{He:2021eiu} could now be extended to
these cases, and to glean interesting organizing principles by doing so. The pictorial representation of putative CFTs obtained by PCA, shown in Figure \ref{fig:pca} is already suggestive in this direction.

Finally, we note that symmetry in datasets was also characterized using neural networks in \cite{Craven:2021ems,Krippendorf:2020gny,Dillon:2021gag,Barenboim:2021vzh,Desai:2021wbb}. It would be interesting to study how those methods apply to the examples studied here. More generally,  we also note that ideas and methods from mathematical physics, in particular quantum field theory, have been used to explicate the 
dynamics of neural networks (see e.g. \cite{dmk,Halmagyi:2020vck,Halverson:2020trp,Erbin:2021kqf}). Indeed, the very structures of quantum field theory and holography appear to be closely connected to certain neural networks \cite{Hashimoto:2019bih}. 
We therefore expect that the interplay between mathematics, physics and data science will yield many new insights of mutual interest and benefit over the long term as well.
\paragraph{Acknowledgements:} Figures \ref{fig:toy4pt} and \ref{fig:pca} are reproduced from the originals in \cite{Chen:2020dxg}. SL's research was supported by
the Simons Collaboration on the Non-perturbative Bootstrap and the Faculty of Sciences, University of Porto while \cite{Chen:2020dxg,Chen:2020jjw} were carried out.

\end{document}